\documentclass[aps,prd,onecolumn,groupedaddress,showpacs,nofootinbib,amssymb]{revtex4}
\usepackage[dvips]{graphicx}
\usepackage{amssymb}
\usepackage{amsmath}
\usepackage{graphicx,,color}
\usepackage{amsfonts}
\usepackage{bm}
\usepackage{cancel}
\usepackage{comment}

\newcommand\be{\begin{equation}}
\newcommand\ee{\end{equation}}

\allowdisplaybreaks[4]

\begin{document}

\tolerance=5000

\title{$f(R)$ Gravity $k$-Essence Late-time Phenomenology}
\author{S.D.~Odintsov,$^{1,2,3}$\,\thanks{odintsov@ieec.uab.es}
V.K.~Oikonomou,$^{4,5,3}$\,\thanks{v.k.oikonomou1979@gmail.com}F.P.
Fronimos,$^{4}$\,\thanks{fotisfronimos@gmail.com}}
\affiliation{$^{1)}$ ICREA, Passeig Luis Companys, 23, 08010 Barcelona, Spain\\
$^{2)}$ Institute of Space Sciences (IEEC-CSIC) C. Can Magrans
s/n,
08193 Barcelona, Spain\\
$^{3)}$ Tomsk State Pedagogical University, 634061 Tomsk, Russia\\
$^{4)}$ Department of Physics, Aristotle University of
Thessaloniki, Thessaloniki 54124,
Greece\\
$^{5)}$ Laboratory for Theoretical Cosmology, Tomsk State
University of Control Systems and Radioelectronics, 634050 Tomsk,
Russia (TUSUR)}

\tolerance=5000

\begin{abstract}
In this work we shall study the late-time behavior of $k$-Essence
$f(R)$ gravity without scalar potential, in the presence of matter
and radiation perfect fluids. We quantify the late-time study by
using the statefinder function
$Y_H(z)=\frac{\rho_{DE}}{\rho_m^{(0)}}$, which is a function of
the redshift and of the Hubble rate. By appropriately rewriting
the Friedmann equation in terms of the redshift and of the
function $Y_H(z)$, we numerically solve it using appropriate
initial conditions, and we critically examine the effects of the
$k$-Essence higher order kinetic terms. As we demonstrate, the
effect of the higher order scalar field kinetic terms on the
late-time dynamics is radical, since the dark energy oscillations
are absent, and in addition, the cosmological physical quantities
are compatible with the latest Planck data and also the model is
almost indistinguishable from the $\Lambda$ Cold Dark Matter
model. This is in contrast to the standard $f(R)$ gravity case,
where the oscillations are present. Furthermore, by choosing a
different set of values of two of the free parameters of the
model, and specifically the coefficient of the higher order
kinetic term and of the exponent of $R^{\delta}$ appearing in the
$f(R)$ action, we demonstrate that it is possible to obtain
$\rho_{DE}<0$ for redshifts $z\sim 2-3.8$, which complies
phenomenologically with, and seems to explain, the observational
data for the same redshifts, and also to obtain a viable
cosmological evolution at $z\sim 0$, at least when the dark energy
equation of state parameter and the dark energy density parameters
are considered.
\end{abstract}


\maketitle

\section{Introduction}

The last 25 years were crucial for the development of cosmology,
since the observations elevated theoretical cosmology to be what
we now know as precision cosmology. The observational data coming
from the cosmic background radiation (CMB) confirmed a nearly
scale invariant power spectrum of primordial curvature
perturbations, while the observational data coming from SNIa
standard candles indicated a striking late-time phenomenon, the
currently accelerating Universe \cite{Riess:1998cb}. The late-time
era, is usually dubbed dark energy era, due to the fact that the
physical process that drives this late-time acceleration era is
still unknown. In order to have acceleration in standard
Einstein-Hilbert gravity, the equation of state (EoS) parameter of
the fluid that drives the acceleration must be $w<-\frac{1}{3}$,
and a negative pressure is the main characteristic of the fluid
that drives late-time acceleration. The cosmological constant
$\Lambda $, is the simplest quantity that may generate the
late-time acceleration, and up to date, the so-called $\Lambda$
Cold Dark Matter ($\Lambda$CDM) model is the most successful
description of late-time physics, being quite compatible with the
CMB data.

Apart from the successes of the $\Lambda$CDM model, there are
several questions unanswered, mainly having to do with the
dynamical nature of dark energy. In the $\Lambda$CDM model, the
EoS parameter of dark energy is constant, however although this is
compatible with the observations, it is not certain that the EoS
parameter is constant. In fact, it might be evolving from a
quintessential value to a phantom value. Apart from the above
issue, the $H_0$-tension
\cite{Aylor:2018drw,Wong:2019kwg,Verde:2019ivm,Knox:2019rjx,Riess:2016jrr,Migkas:2017vir,Ramos-Ceja:2019zxt}
turns out to be a serious troubling problem that needs to be
explained in a theoretical way, and has recently been discussed in
the literature
\cite{Doran:2006kp,Bhattacharyya:2019lvg,Sakstein:2019fmf,Tian:2019enx,Nojiri:2019fft}.
Another quite important issue is the discrepancy between the CMB
based value of $\Omega_m h^2\simeq 0.12\pm 0.001$ and the one
which is evaluated from the Friedmann equation at $z=2.34$, if one
substitutes the observed value $H(z=2.34)=222 \pm 7 Km/Mpc/s$,
which indicates that $\Omega_m h^2\simeq 0.132\pm 0.008$ if dark
energy is absent \cite{Delubac:2014aqe,Sahni:2014ooa}. The
description given in Ref. \cite{Sahni:2014ooa} perfectly describes
this issue, so we now share the description of Ref.
\cite{Sahni:2014ooa} in order to clarify how the findings of
\cite{Delubac:2014aqe} indicate a possible tension with the value
of $\Omega_m h^2$ obtained from the CMB or the $\Lambda$CDM model.
If the general relativistic Friedmann equation is assumed,
$H^{2}(z)=\frac{\kappa^2}{3}\left( \rho_m^0(1+z)^3 \right)$
($\kappa^2=8\pi G$), or equivalently,
$\frac{h^2(z)}{(1+z)^3}=\Omega_{m}h^2$, then by substituting
$H(z=2.34)=222 \pm 7 Km/Mpc/s$, then we obtain
$\frac{h^2(z)}{(1+z)^3}\simeq 0.132\pm 0.008$, which is different
from the CMB value $\Omega_m h^2\simeq 0.12\pm 0.001$. This is a
serious issue, which should be theoretically explained.

It is conceivable that such theoretical issues cannot be harbored
by standard Einstein-Hilbert gravity, and require formal
extensions that may describe such involved physical behaviors.
Modified gravity provides a solid theoretical framework in the
context of which phenomena such as the dark energy era, and also
the dark matter issue can be consistently be described accurately,
for reviews see
\cite{Nojiri:2017ncd,Nojiri:2009kx,Capozziello:2011et,Capozziello:2010zz,Nojiri:2006ri,
Nojiri:2010wj,delaCruzDombriz:2012xy,Olmo:2011uz}, however with
regard to dark matter, the particle dark matter description
\cite{Bertone:2004pz,Bergstrom:2000pn,Mambrini:2015sia,Profumo:2013yn,Hooper:2007qk,Oikonomou:2006mh}
still seems to be supported from observations, like in the bullet
cluster. In fact, it is possible to describe in a unified way both
the inflationary era and the late-time acceleration eras, using
the same theoretical framework. This was firstly demonstrated in
the context of $f(R)$ gravity in Ref. \cite{Nojiri:2003ft}, and
several other $f(R)$ gravity unified cosmologies appeared in the
literature
\cite{Nojiri:2007as,Nojiri:2007cq,Cognola:2007zu,Nojiri:2006gh,Appleby:2007vb,Elizalde:2010ts,Odintsov:2020nwm}.
However, although $f(R)$ gravity can describe a viable late-time
era, compatible with the observational data and the $\Lambda$CDM
model, there is a feature that haunts the $f(R)$ gravity
description of the dark energy era, namely the dark energy
oscillations at larger redshifts \cite{Bamba:2012qi}.
Particularly, it is known that, due to the presence of higher
derivatives of the Hubble rate, the $f(R)$ gravity description of
late-time evolution is plagued with dark energy oscillations
during the last stages of the matter domination
\cite{Bamba:2012qi}. These oscillations are even more enhanced if
statefinder quantities are considered, such as the deceleration
parameter and the jerk. In general if quantities that contain
higher derivatives of the Hubble rate are considered, the
oscillations are more pronounced.

In this work we shall study the late-time behavior of $k$-Essence
$f(R)$ gravity models, with the $k$-Essence part containing only a
canonical kinetic term for the scalar field, and higher order
kinetic terms, without the presence of a scalar potential.
$k$-Essence theories themselves are quite interesting
phenomenologically, since firstly these survived after the
striking GW170817 event \cite{GBM:2017lvd} in 2017 (see Ref.
\cite{Ezquiaga:2017ekz} for a complete list of the viable modified
gravity theories), which indicated that the gravitational wave
speed is equal to one in natural units. Apart from this important
feature, $k$-Essence theories can describe in a viable way both
inflation and the late-time era, and for an important stream of
papers on this issue see
\cite{ArmendarizPicon:1999rj,Chiba:1999ka,ArmendarizPicon:2000dh,Matsumoto:2010uv,ArmendarizPicon:2000ah,Chiba:2002mw,Malquarti:2003nn,Malquarti:2003hn,Chimento:2003zf,Chimento:2003ta,Scherrer:2004au,Aguirregabiria:2004te,ArmendarizPicon:2005nz,Abramo:2005be,Rendall:2005fv,Bruneton:2006gf,dePutter:2007ny,Babichev:2007dw,Deffayet:2011gz,Kan:2018odq,Unnikrishnan:2012zu,Li:2012vta,Gialamas:2019nly,Nojiri:2019dqc,Odintsov:2019ahz}.
For the purposes of this work, we shall choose an appropriate
$f(R)$ gravity, in the presence of dust and radiation perfect
fluids, which is extensively studied in Ref.
\cite{Odintsov:2020nwm}, see also \cite{Nojiri:2019fft}, which can
describe in a unified way both the inflationary era and the dark
energy era, and also can describe an early dark energy era, by
adding an appropriate early dark energy term. We shall call the
$f(R)$ gravity model of Ref. \cite{Odintsov:2020nwm}, power-law
corrected $R^2$ model, just for the purposes of this paper, in
order to discriminate it from other power-law $f(R)$ gravity
models which contain powers of the curvature. As it is shown in
Ref. \cite{Odintsov:2020nwm}, the power-law corrected $R^2$ model
produces a viable late-time phenomenology, compatible with the
Planck 2018 data \cite{Aghanim:2018eyx}, and mimics to a great
extent the $\Lambda$CDM model \cite{Odintsov:2020nwm}. We shall
incorporate to the theory the $k$-Essence terms, and by
numerically solving\footnote{We used Mathematica 9$^{\circledR}$}
the Friedmann equation, we shall explore the effects of the
$k$-Essence terms on the $f(R)$ gravity late-time phenomenology.
For our study, we shall express all the physical quantities in
terms of the statefinder function
$Y_H(z)=\frac{\rho_{DE}}{\rho_m^{(0)}}$, which is a function of
the redshift and of the Hubble rate. As we demonstrate, for a
specific set of values of the free parameters of the model, the
dark energy oscillations at large redshifts of the order $z\sim
10$, which are present in the simple $f(R)$ gravity model, are
absent in the case of the $k$-Essence $f(R)$ gravity model, while
at the same time the cosmological evolution remains viable and
compatible with the Planck 2018 observational data and the
$\Lambda$CDM model. Moreover, by using another set of values of
the free parameters, we show that it is possible to comply with
the observations of \cite{Delubac:2014aqe} on the value of the
Hubble rate at $z\sim 2.34$. Our results indicate that the
$k$-Essence terms may actually act as a compensating dark energy
mechanism of the $f(R)$ gravity effective fluid, and at the same
time a viable evolution at $z\sim 0$ is obtained. For our analysis
we investigate the behavior of several well-known statefinder
functions, and we compare the results with the $\Lambda$CDM values
and with the observational data.

This paper is organized as follows: In section II we present and
discuss the theoretical model of $k$-Essence $f(R)$ gravity. In
section III we introduce the function $Y_H(z)$ and by expressing
the physical quantities in terms of $Y_H(z)$ and the redshift, we
rewrite the Friedmann equation in terms of $Y_H(z)$ and its
derivatives. In addition, in section III, we study numerically the
late-time behavior of a specific $k$-Essence $f(R)$ gravity model
and we compare the results to the power-law $f(R)$ gravity model
and the $\Lambda$CDM model. Accordingly, we demonstrate how the
$k$-Essence $f(R)$ gravity model can explain the 2014 results on
the Hubble rate value for redshifts $z\sim 2.34$ without the need
for introducing a compensating dark energy term. Finally, the
conclusions of our work follow at the end of the paper.

\section{$f(R)$ Gravity $k$-Essence Framework}

The $k$-Essence $f(R)$ gravity theory belongs to the general class
of theories of the form $f(R,X,\phi)$, with $X=\frac{1}{2}
\partial_{\mu}\phi \partial^{\mu}\phi $. We shall assume that the
gravitational action is,
\begin{equation}\label{ft1}
\mathcal{S}=\int
d^4x\sqrt{-g}\Big{(}\frac{f(R)}{2\kappa^2}+G(X)+\mathcal{L}_{matter}\Big{)}\,
,
\end{equation}
where $f(R)$ is an arbitrary function of the Ricci scalar to be
specified later on, $G(X)$ is a function depending solely on the
kinetic term $X=\frac{1}{2} \partial_{\mu}\phi \partial^{\mu}\phi
$ and $\kappa^2=8\pi G=\frac{1}{M_p^2}$, where $G$ is Newton's
constant and $M_p$ is the reduced Planck mass. In addition,
$\mathcal{L}_{matter}$ denotes the Lagrangian of the perfect
matter fluids that are present. Moreover, the background geometry
will be assumed to be a flat Friedmann-Robertson-Walker (FRW)
metric, with line element,
\begin{equation}
\label{metricfrw} ds^2 = - dt^2 + a(t)^2 \sum_{i=1,2,3}
\left(dx^i\right)^2\, ,
\end{equation}
where $a(t)$ is the scale factor. From now on, we assume that the
scalar field is homogenous, meaning it is only time dependent.
Recalling the definition of the kinetic term $X$ and the line
element, we get,
\begin{equation}\label{ft3}
X=-\frac{1}{2}\dot\phi^2\, ,
\end{equation}
In order to find the equations of motion, we vary the
gravitational action (\ref{ft1}) with respect to the metric tensor
and to the scalar field, and the gravitational equations of motion
are,
\begin{equation}\label{ft4}
\begin{aligned}
\kappa^2\rho+\frac{1}{2}(FR-f)+\kappa^2G_X(X)X-3H\dot F=3FH^2\, ,\\
\kappa^2(\rho+P)+\ddot F-H\dot F+2\dot H F+\kappa^2G_X(X)X=0\, ,\\
\frac{1}{a^3}\frac{d}{dt}(a^3G_X(X)\dot\phi)=0 \, ,
\end{aligned}
\end{equation}
where $F=\frac{\partial f}{\partial R}$, $G_X=\frac{\partial
G}{\partial X}$, and $\rho$ stands for the energy density of the
matter perfect fluids that are present, and $P$ is the
corresponding pressure. Also

For the purposes of this work we shall assume that both
non-relativistic matter (cold dark matter and baryons) and
relativistic matter (radiation) are present, so $\rho$ is equal
to,
\begin{equation}\label{ft5}
\rho=\rho_{m}^{(0)}(\frac{1}{a^3}+\chi\frac{1}{a^4})\, ,
\end{equation}
where $\chi=\frac{\rho_{r}^{(0)}}{\rho_{m}^{(0)}}$. As we already
mentioned in the introduction, the gravitational wave speed (speed
of tensor metric perturbations) is for the $f(R,X,\phi)$ theory at
hand $c_T^2=1$, but we need to mention for the sake of
completeness that the sound wave speed of the perturbations of the
theory is non-trivial,
\begin{equation}\label{wavespeed}
c_A^2=\frac{XG_X+\frac{3\dot{F}^2}{2F}}{XG_X+2X^2G_{XX}+\frac{3\dot{F}^2}{2F}}\,
,
\end{equation}
with $G_{XX}=\frac{\partial^2 G}{\partial X^2}$. However, this
wave speed affects the scalar and tensor perturbations, and will
not affect the late-time behavior of the model. Having presented
in brief the theoretical framework we shall consider, in the next
section we shall express the gravitational equations in terms of
suitable statefinder functions and in terms of the redshift and we
shall consider the late-time behavior of a specific $f(R,X,\phi)$
model, with quite interesting late-time phenomenology.

\section{A Viable $f(R)$ Gravity $k$-Essence Model and Comparison with Standard $f(R)$ Gravity}

In order to study the late-time era of the $k$-Essence $f(R)$
gravity model, we shall introduce appropriate functions that will
quantify our study accurately. Firstly, we shall use the redshift
as a dynamical variable, defined as,
\begin{equation}\label{ft6}
1+z=\frac{1}{a}\, ,
\end{equation}
where we took that the present scale factor of the Universe is
unity, so present time corresponds to $z=0$. By using,
\begin{equation}\label{ft7}
\frac{d}{dt}=-H(1+z)\frac{d}{dz}\, ,
\end{equation}
and $H=H(z)$ we shall express all the quantities in the
gravitational equations as functions of the redshift. The
derivatives with respect to the cosmic time, correspond to the
following derivatives with respect to the redshift,
\begin{equation}\label{ft8}
\begin{aligned}
\dot F=-H(1+z)F_{z}\, ,\\
\dot H=-H(1+z)H_{z}\, ,\\
\dot\phi=-H(1+z)\phi_{z}\, ,\\
\ddot F=H^2(1+z)^2F_{zz}+(1+z)H^2F_{z}+HH_{z}(1+z)^2F_{z}\, ,\\
\end{aligned}
\end{equation}
where $F_z=\frac{d F}{d z}$ and $F_{zz}=\frac{d^2 F}{d z^2}$. We
shall use this notation hereafter, so the subscript to a function
will mean the total or partial derivative of the function with
respect to the variable appearing to the subscript. The Ricci
scalar for a flat FRW spacetime is,
\begin{equation}\label{ft9}
R=12H^2+6\dot H\, ,
\end{equation}
so this can be expressed as a function of the redshift as follows,
\begin{equation}\label{ft10}
R=12H^2-6HH_{z}(1+z)\, .
\end{equation}
Also, the equation of motion of the scalar field is easily
obtained,
\begin{equation}\label{ft11}
H(\frac{dG_X}{dz}H\phi_{z}+G_XH_{z}\phi_{z}+G_XH\phi_{zz}-\frac{2G_XH\phi_{z}}{1+z})=0\,
.
\end{equation}
In order to better quantify the late-time behavior of the
$k$-Essence $f(R)$ gravity model, we shall introduce the following
function $Y_H(z)$ \cite{Hu:2007nk,Bamba:2012qi},
\begin{equation}\label{yHdefinition}
Y_H(z)=\frac{\rho_{DE}}{\rho^{(0)}_m}\, ,
\end{equation}
with $\rho^{(0)}_m$ being the present time energy density of
non-relativistic matter. In the above equation, $\rho_{DE}$ is the
energy density of the dark energy fluid, which now consists of the
$k$-Essence terms and of the $f(R)$ gravity terms. Actually, the
field equations can be cast in an Einstein-Hilbert form for a flat
FRW metric as follows,
\begin{align}\label{flat}
& 3H^2=\kappa^2\rho_{tot}\, ,\\ \notag &
-2\dot{H}=\kappa^2(\rho_{tot}+P_{tot})\, ,
\end{align}
with $\rho_{tot}=\rho_{m}+\rho_{DE}+\rho_r$ denoting the total
energy density of the effective cosmological fluid and
correspondingly, $P_{tot}=P_r+P_{DE}$ stands for the total
pressure of the cosmological fluid. The effective cosmological
fluid in our case receives contributions from the cold dark matter
($\rho_m$) and radiation fluids ($\rho_r$), but also from the
combined $k$-Essence and $f(R)$ gravity fluids ($\rho_{DE}$), with
the latter fluids being responsible for the late-time evolution.
The energy density of dark energy fluid, which has combined
contributions from the $k$-Essence and $f(R)$ gravity fluids, is
equal to,
\begin{equation}\label{degeometricfluid}
\kappa^2\rho_{DE}=\frac{1}{2}(FR-f)+\kappa^2G_X(X)X-3H\dot
F-3H^2(F-1)\, ,
\end{equation}
which can easily be read off from the Friedmann equation
(\ref{ft4}). Accordingly, from the Raychaudhuri equation
(\ref{ft4}), the pressure of the dark energy fluid is,
\begin{equation}\label{pressuregeometry}
\kappa^2P_{DE}=\ddot F-H\dot
F+2\dot{H}(F-1)+\kappa^2G_X(X)X-\kappa^2\rho_{DE}\, .
\end{equation}
In the way chosen in the above equations, all the fluids that
constitute the total cosmological effective fluid, are perfect
fluids, non-interacting, and satisfy the continuity equations,
\begin{align}\label{fluidcontinuityequations}
& \dot{\rho}_a+3H(\rho_a+P_a)=0\, , \\ \notag &
\dot{\rho}_r+3H(\rho_r+P_r)=0\, , \\ \notag &
\dot{\rho}_{DE}+3H(\rho_{DE}+P_{DE})=0\, .
\end{align}
Having the fluid descriptions at hand, the function $Y_H(z)$
defined in Eq. (\ref{yHdefinition}) can be written in terms of the
Hubble rate by using the Friedman equation,
\begin{equation}\label{akyri}
3H^2=\kappa^2 (\rho(z)+\rho_{DE})\, ,
\end{equation}
and it reads,
\begin{equation}\label{finalexpressionyHz}
Y_H(z)=\frac{H^2}{m_s^2}-(1+z)^{3}-\chi (1+z)^4\, .
\end{equation}
Obviously, the function $Y_H(z)$ is a statefinder function since
it depends only on the Hubble rate and the redshift. Also the
parameter $m_s^2$ appearing in Eq. (\ref{finalexpressionyHz}) is
$m_s^2=\frac{\kappa^2\rho^{(0)}_m}{3}=H_0\Omega_m=1.87101\times
10^{-67}$eV$^2$, and we used the current observational data coming
from Planck 2018 \cite{Aghanim:2018eyx} for the definition of the
Hubble rate and $\Omega_m$ (see also later on in this section the
discussion on the values of the cosmological parameters). We shall
express every quantity entering the Friedmann equation in terms of
the function $Y_H(z)$, so practically, the Hubble rate is
expressed in terms of the function $Y_H(z)$, and we have,
\begin{equation}\label{neweqnv1}
H^2=m_s^2(Y_H+\frac{\rho}{\rho_{m}^{(0)}})\, ,
\end{equation}
and accordingly, by differentiating the above with respect to $z$
we get,
\begin{equation}\label{neweqnv2}
HH_{z}=\frac{m_s^2}{2}(\frac{dY_H}{dz}+\frac{\rho_{z}}{\rho_{m}^{(0)}})\,
,
\end{equation}
where the subscript ``$z$'' denotes differentiation with respect
to z. Assuming that the $f(R)$ gravity function is written as
$f(R)=R+f_0(R)$ where again, $f_0(R)$ is an arbitrary function of
the Ricci scalar, then $F=\frac{d f}{d R}$ reads,
\begin{equation}\label{ft13}
F=1+\frac{d f_0}{d R}\, .
\end{equation}
Similarly, the derivative of $F$ with respect to $z$ is equal to,
\begin{equation}\label{ft14}
F_{z}=\frac{d}{dz}(\frac{df_0}{dR})\, .
\end{equation}
It is vital to derive an expression for the derivative of the
Ricci scalar with respect to redshift $z$, so we have,
\begin{equation}\label{ft15}
\frac{dR}{dz}=18HH_{z}-6(1+z)(H_{z}^2+HH_{zz})\, ,
\end{equation}
In addition, by further differentiating Eq. (\ref{neweqnv2}) we
get,
\begin{equation}\label{ft16}
H_{z}^2+HH_{zz}=\frac{m^2}{2}(\frac{d^2Y_H}{dz^2}+\frac{\rho_{zz}}{\rho_{m}^{(0)}})\,
.
\end{equation}
Finally, another useful expression is the derivative of the
kinetic term $X$ with respect to the redshift $z$, which is,
\begin{equation}\label{ft19}
\frac{dX}{dz}=-(1+z)\phi_{z}((1+z)HH_{z}\phi_{z}+H^2((1+z)\phi_{zz}+\phi_{z}))\,
.
\end{equation}
Thus the gravitational equations that we will solve numerically
have the following form,
\begin{align}\label{ft20}
&\kappa^2\rho+\frac{1}{2}(FR-f)+\kappa^2G_X(X)X+3H^2\Big{(}(1+z)F_{z}-F\Big{)}=0\,
, \\ \notag &
H\Big{(}\frac{dG_X}{dz}H\phi_{z}+G_XH_{z}\phi_{z}+G_XH\phi_{zz}-\frac{2G_XH\phi_{z}}{1+z}\Big{)}=0\,
,
\end{align}
and the following definitions and expressions shall be used,
firstly $\rho(z)$ as a function of the redshift,
\begin{equation}\label{ext1}
\rho=\rho_{m}^{(0)}((1+z)^{3}+\frac{\rho_{r}^{(0)}}{\rho_{m}^{(0)}}(1+z)^{4})=\rho_{m}^{(0)}((1+z)^{3}+\chi(1+z)^{4})\,
,
\end{equation}
and then, the Hubble rate and its derivatives with respect to the
redshift,
\begin{align}\label{ext2}
& H^2=m_s^2(Y_H+\frac{\rho}{\rho_{m}^{(0)}})\, ,\\ \notag &
HH_{z}=\frac{m_s^2}{2}(\frac{dY_H}{dz}+\frac{\rho_{z}}{\rho_{m}^{(0)}})\,
,\\ \notag &
H_{z}^2+HH_{zz}=\frac{m_s^2}{2}(\frac{d^2Y_H}{dz^2}+\frac{\rho_{zz}}{\rho_{m}^{(0)}})\,
.
\end{align}
Furthermore, the Ricci scalar, its derivative with respect to the
redshift and $F$ and $F_z$ are,
\begin{align}\label{ext3}
& R=12H^2-6HH_{z}(1+z)\, ,\\ \notag &
\frac{dR}{dz}=18HH_{z}-6(1+z)(H_{z}^2+HH_{zz})\, ,\\ \notag &\
F=1+\frac{df_{0}}{dR}\, ,\\ \notag &
Fz=\frac{d^2f_{0}}{dR^2}\frac{dR}{dz}\, .
\end{align}
At this point, we shall specify the $G(X)$ function appearing in
Eq. (\ref{ft1}), so we assume that,
\begin{equation}\label{ft17}
G(X)=\beta(X+\frac{1}{2}f_{1}X^{m})\, ,
\end{equation}
where $\beta$ is a dimensionless parameter which will be set equal
to $\beta=-1$ in order to have a canonical kinetic term.  We chose
to leave this in general form in $G(X)$, and not equal to $-1$, in
order to have the phantom scalar case available, but this is not
our case though. Also $f_1$ has mass dimensions $[m]^{4-m}$. For
the model (\ref{ft17}) we have,
\begin{align}\label{ft18}
& G_X(X)=\beta(1+\frac{m}{2}f_{1}X^{m-1})\, , \\ \notag & \frac{d
G_X(X)}{dz}=\beta\frac{m(m-1)}{2}f_{1}X^{m-2}\frac{dX}{dz}\, .
\end{align}
with the derivative of the kinetic term with respect to the
redshift being equal to,
\begin{equation}\label{derivkineticx}
\frac{dX}{dz}=-(1+z)\phi_{z}((1+z)HH_{z}\phi_{z}+H^2((1+z)\phi_{zz}+\phi_{z}))\,
,
\end{equation}
and the functions $H$ and $H_z$ are given in terms of the function
$Y_H(z)$ in Eqs. (\ref{ext2}).
\begin{figure}[h!]
\centering
\includegraphics[width=20pc]{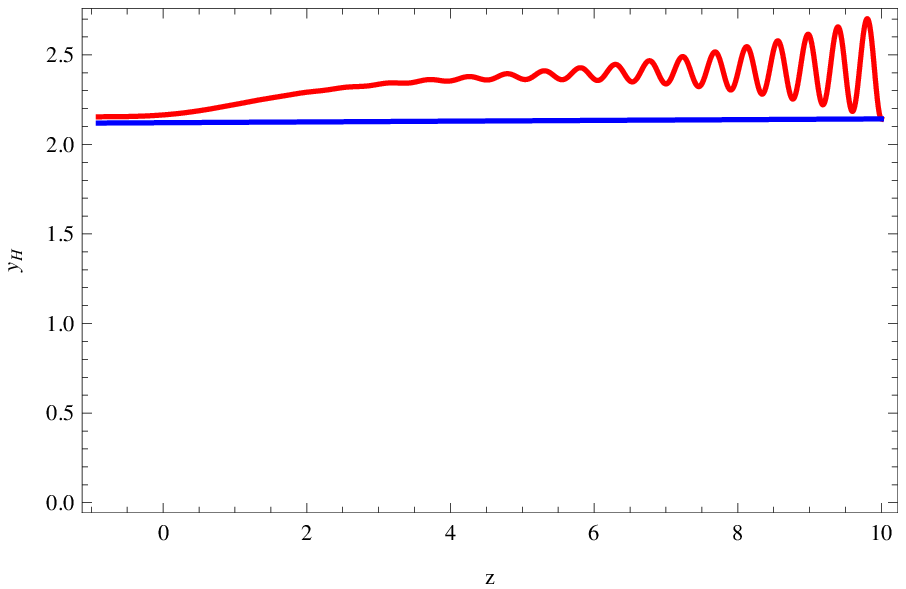}
\includegraphics[width=20pc]{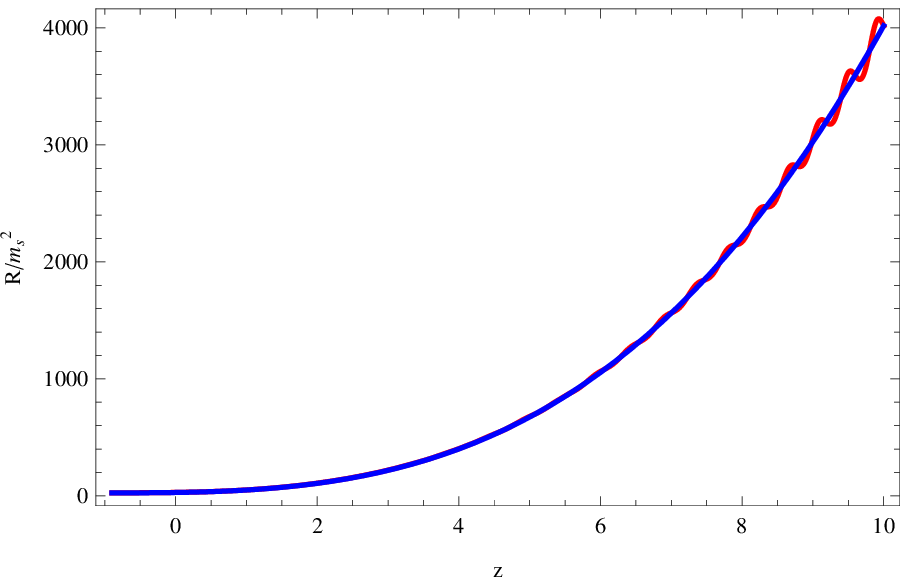}
\caption{Plots of the statefinder function $Y_H$ (left plot), and
of  $R/m_s^2$ (right plot) for the $k$-Essence $f(R)$ gravity
(blue curves) and for the power-law corrected $R^2$ model (red
curves) as functions of the redshift.} \label{plot1}
\end{figure}

\subsection{Late-time $k$-Essence $f(R)$ Gravity Dynamics}

At this point, let us specify the $f(R)$ gravity function in order
to quantify the effect of the $k$-Essence terms on the late-time
dynamics of the $f(R)$ gravity theory. We shall choose the
following $f(R)$ gravity \cite{Odintsov:2020nwm},
\begin{equation}\label{starobinsky}
f(R)=R+\frac{1}{M^2}R^2-\gamma \Lambda
\Big{(}\frac{R}{3m_s^2}\Big{)}^{\delta}\, .
\end{equation}
In the above equation, the parameter $m_s^2$ was defined below Eq.
(\ref{finalexpressionyHz}), and also $\delta$ is freely chosen in
the interval $0<\delta <1$, while $\gamma$ is equal to $\gamma=2$.
The parameter $\delta$ shall be chosen equal to $\delta=1/100$ for
late-time phenomenological reasons \cite{Odintsov:2020nwm}.
Moreover, the value of the parameter $\Lambda$ will be given later
on in this section. Also, the parameter $M$ is chosen for
inflationary phenomenological reasons equal to $M= 1.5\times
10^{-5}\left(\frac{N}{50}\right)^{-1}M_p$ \cite{Appleby:2009uf},
with $N$ being the $e$-foldings number during the inflationary
era. The phenomenology of the model (\ref{starobinsky}) is
thoroughly investigated in Ref. \cite{Odintsov:2020nwm} both at
early and late-times. As is shown in \cite{Odintsov:2020nwm}, the
model (\ref{starobinsky}) can generate a successful inflationary
era, due to the presence of the $R^2$ term, and at the same time
can produce a viable dark energy era, compatible to the
observational data of Planck 2018, and also very similar to the
$\Lambda$CDM model. The details of this investigation can be found
in Ref. \cite{Odintsov:2020nwm}. Furthermore, it is shown in Ref.
\cite{Nojiri:2019fft} that if an appropriate term is added in the
$f(R)$ gravity (\ref{starobinsky}), an early dark energy era can
also be realized by the same model. Also, the presence of the
$R^2$ term is vital for the disappearance of singularities in the
dark energy EoS during the matter domination era
\cite{Appleby:2009uf,Bamba:2008ut}. Our aim in this work is to
study the effects of the $k$-Essence terms in the late-time
phenomenology of the $f(R)$ gravity of Eq. (\ref{starobinsky}). As
is shown in Ref. \cite{Nojiri:2019fft}, the late-time era of the
$f(R)$ gravity model (\ref{starobinsky}) is plagued by dark energy
oscillations, which are more enhanced when statefinder functions
that contain higher derivatives of the Hubble rate are considered.
As we show, the dark energy oscillations are significantly damped
due to the presence of the $k$-Essence terms, if appropriate
initial conditions are chosen for the scalar field.
\begin{figure}[h!]
\centering
\includegraphics[width=20pc]{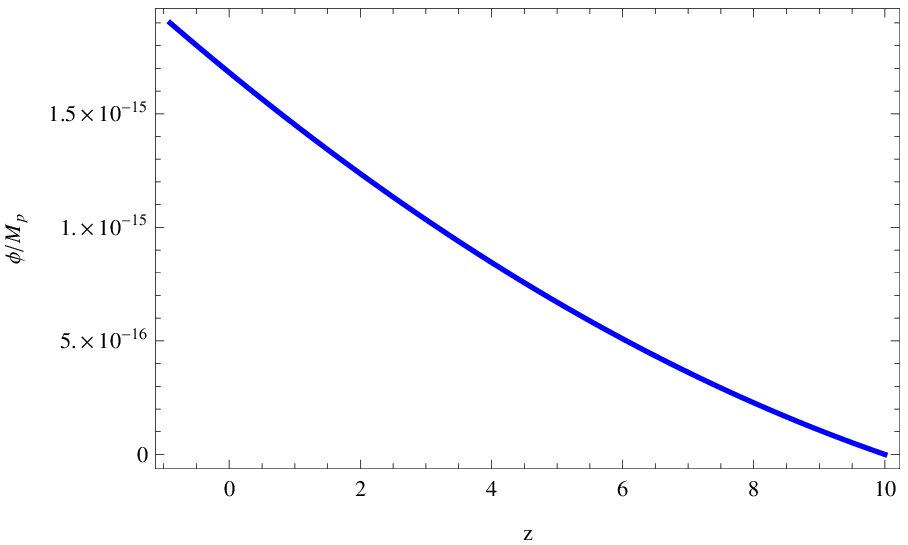}
\includegraphics[width=20pc]{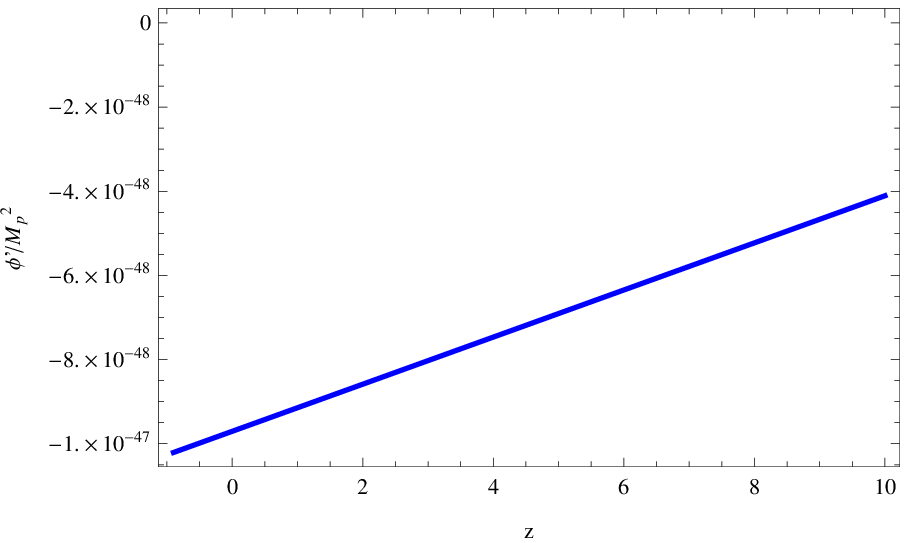}
\caption{Plots of  $\phi (z)$ (left plot) and of $\phi'(z)$ (right
plot) for the $k$-Essence $f(R)$ gravity as functions of the
redshift.} \label{plot2}
\end{figure}
With regard to the values of the cosmological parameters, we shall
assume that the Hubble rate is \cite{Aghanim:2018eyx},
\begin{equation}\label{H0today}
H_0=67.4\pm 0.5 \frac{km}{sec\times Mpc}\, ,
\end{equation}
so $H_0=67.4km/sec/Mpc$ or equivalently $H_0=1.37187\times
10^{-33}$eV, therefore $h\simeq 0.67$. Hence, we shall take into
account only the CMB based value of the Hubble rate only,
disregarding the Cepheid based value. In addition, according to
the CMB extracted observational data $\Omega_c h^2$ is,
\begin{equation}\label{codrdarkmatter}
\Omega_c h^2=0.12\pm 0.001\, ,
\end{equation}
which we also used earlier for the definition of the parameter
$m_s^2$. Furthermore the parameter $\Lambda$ appearing in Eq.
(\ref{starobinsky}), will be taken equal to $\Lambda\simeq
11.895\times 10^{-67}$eV$^2$ and in addition, the parameter
$m_s^2$ expressed in eV $m_s^2\simeq 1.87101\times
10^{-67}$eV$^2$, while $M$ is $M\simeq 3.04375\times 10^{22}$eV
for $N\sim 60$. Also, the fraction of the present time radiation
to dark matter energy densities is,
\begin{equation}\label{ft23}
\frac{\rho_{r}^{(0)}}{\rho_{m}^{(0)}}=\chi=3.1\times 10^{-4}\, .
\end{equation}
Furthermore, the parameter denoted as $f_{1}$ appearing in the
$G(X)$ function (\ref{ft17}) of the $k$-Essence part of the
Lagrangian shall be chosen proportional to the parameter
$\Lambda$,
\begin{equation}\label{ft26}
f_{1}\sim \Lambda^{2-2m}\, ,
\end{equation}
where $m$ is the power of the kinetic term in Eq. (\ref{ft17}).
When necessary, the results of this section shall be compared to
the ones corresponding to the $\Lambda$CDM model, so the Hubble
rate of the $\Lambda$CDM model is,
\begin{equation}\label{lambdacdmhubblerate}
H_{\Lambda}(z)=H_0\sqrt{\Omega_{\Lambda}+\Omega_m(z+1)^3+\Omega_r(1+z)^4}\,
,
\end{equation}
with $H_0$ the present value of the Hubble rate, while
$\Omega_{\Lambda}\simeq 0.681369$ and $\Omega_m\sim 0.3153$
\cite{Aghanim:2018eyx}. In addition, while
$\Omega_r/\Omega_m\simeq \chi$, and we defined the parameter
$\chi$ below Eq. (\ref{ft23}).

Let us proceed to the choice of the initial conditions. Our
numerical analysis will be focused on the redshift interval
$z=[0,10]$, and in the following the final redshift value will be
$z_f=10$. For the function $Y_H(z)$ the initial conditions are
chosen to be \cite{Odintsov:2020nwm,Bamba:2012qi},
\begin{align}\label{ft27}
& Y_H(z=z_{f})=\frac{\Lambda}{3m_s^2}(1+\frac{1+z_{f}}{1000})\,
,\\ \notag &
\frac{dY_H}{dz}\Big{|}_{z=z_{f}}=\frac{\Lambda}{3m_s^2}\frac{1}{1000}\,
,
\end{align}
where $z_f$ is the final redshift $z_f=10$. In addition, for the
scalar field the initial conditions are chosen to be,
\begin{equation}\label{initialconditionsscalarfield}
\phi(z=z_{f})= 10^{-20} M_{p}\, ,
\,\,\,\frac{d\phi}{dz}\Big{|}_{z=z_{f}}= -10^{-20} M_{p}\, .
\end{equation}
Also, we shall assume that $f_1$ takes the value $f_1=3\times
10^{-40}\Lambda^{2-2m}$, and also we shall take $m=2$, so we have
a quadratic higher order kinetic term in the Lagrangian of the
$k$-Essence $f(R)$ gravity and also recall that $\beta=-1$ in
order to have a canonical kinetic term for the scalar field. In
effect, the quadratic higher order kinetic term appearing in the
$k$-Essence $f(R)$ gravity is $\sim -f_1 X^2$.

\begin{figure}[h!]
\centering
\includegraphics[width=20pc]{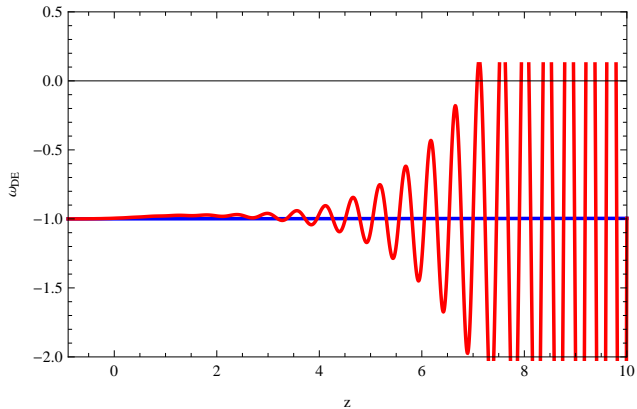}
\caption{Plot of the dark energy EoS parameter $\omega_{DE}(z)$
for the $k$-Essence $f(R)$ gravity (blue curve) and for the
power-law corrected $R^2$ model (red curve) as functions of the
redshift.} \label{plot3}
\end{figure}
\begin{figure}[h!]
\centering
\includegraphics[width=20pc]{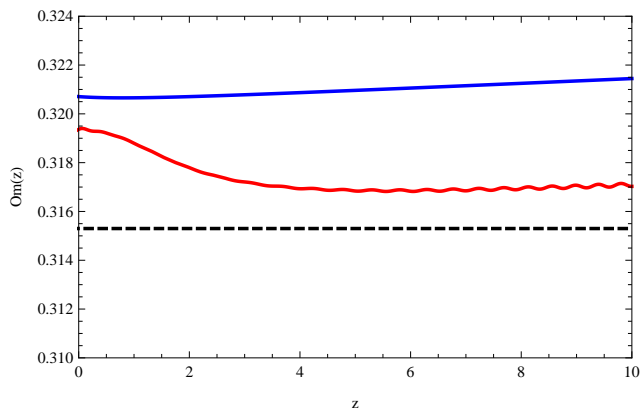}
\caption{Plot of  the statefinder function $Om(z)$ for the
$k$-Essence $f(R)$ gravity (blue curve), for the power-law
corrected $R^2$ model (red curve) and the $\Lambda$CDM model
(black dashed curve) as functions of the redshift.} \label{plot4}
\end{figure}
\begin{figure}[h!]
\centering
\includegraphics[width=20pc]{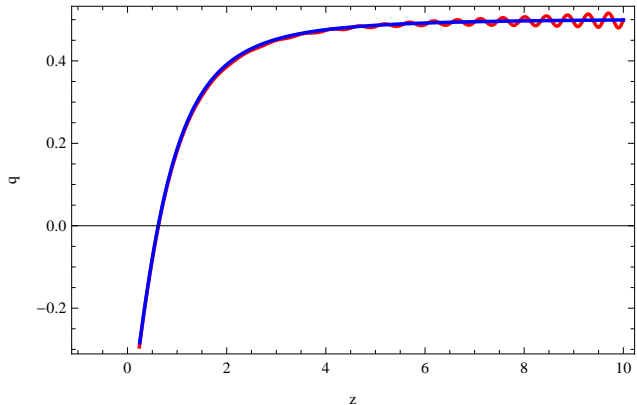}
\caption{Plot of the deceleration parameter $q$ for the
$k$-Essence $f(R)$ gravity (blue curve) and for the power-law
corrected $R^2$ model (red curve) as functions of the redshift.}
\label{plot4a}
\end{figure}
\begin{figure}[h!]
\centering
\includegraphics[width=20pc]{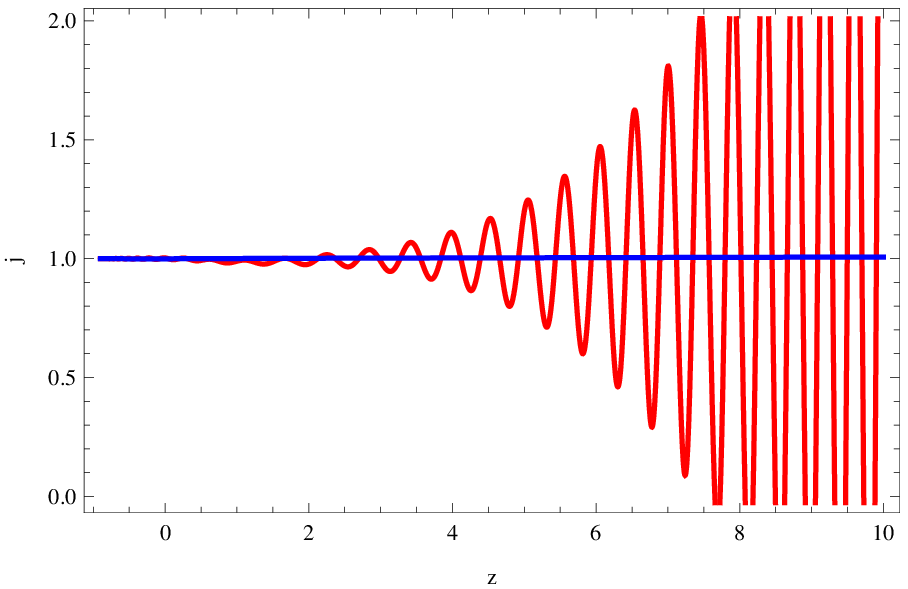}
\includegraphics[width=20pc]{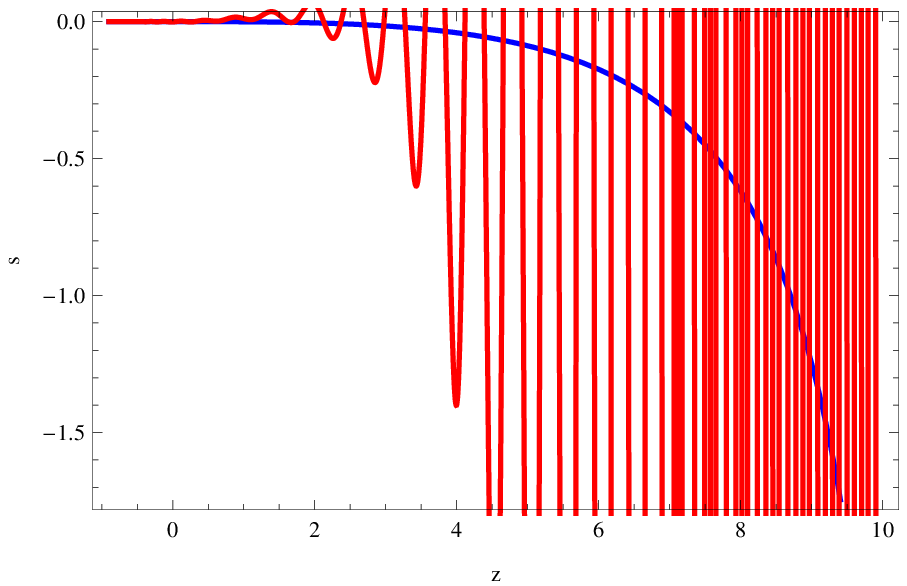}
\caption{Plot of the statefinder function jerk $j$ (left plot) and
the of $s$ (right plot) for the $k$-Essence $f(R)$ gravity (blue
curves), for the power-law corrected $R^2$ model (red curves).}
\label{plot5}
\end{figure}

At this point let us present in detail the results of our
numerical analysis. We shall focus on the behavior of the most
important cosmological quantities, and of the most important
statefinder functions that are used in the literature. Let us
start with the statefinder function $Y_H(z)$ and the curvature
$R$, and in Fig. \ref{plot1} we present the plots of $Y_H$ (left
plot), and of $R/m_s^2$ (right plot) for the $k$-Essence $f(R)$
gravity (blue curves) and for the power-law corrected $R^2$ model
(red curves) as functions of the redshift. As it is obvious in
both the plots, the dark energy oscillations are completely absent
from the $k$-Essence $f(R)$ gravity theory, at least for redshifts
up to $z\sim 10$. Also, it might seem that $Y_H$ for the
$k$-Essence $f(R)$ gravity theory, is constant, however this is
not true, for example at $z=10$ we have $Y_H(10)=2.14249$ while at
redshift $z=0$ we have $Y_H(0)=2.1213$. Also in Fig. \ref{plot2}
we present the plots of $\phi (z)/M_p$ (left plot) and of
$\phi'(z)/M_p^2$ (right plot) for the $k$-Essence $f(R)$ gravity
model as functions of the redshift. As it can be seen, the values
of the scalar field increase as the redshift drops. It is also
notable that if we choose an initial condition $\phi'(z)>0$ at
$z=10$, the scalar field takes negative values, but we did not
study more this case. Let us proceed to the behavior and values of
some important cosmological quantities, starting with the dark
energy EoS parameter $\omega_{DE}=\frac{P_{DE}}{\rho_{DE}}$, which
in terms of $Y_H$ is given below,
\begin{equation}\label{omegade}
\omega_{DE}(z)=-1+\frac{1}{3}(z+1)\frac{1}{Y_H(z)}\frac{d
Y_H(z)}{d z}\, .
\end{equation}
Notably, the dark energy EoS parameter is also a statefinder
quantity since it depends implicitly on $H(z)$ and its higher
first order derivatives. The value of the EoS parameter at present
time is evaluated to be $\omega_{DE}=-0.999667$ for the
$k$-Essence $f(R)$ gravity model, which is compatible with the
latest Planck 2018 data \cite{Aghanim:2018eyx} values
$\omega_{DE}=-1.018\pm 0.031$. Furthermore, in Fig. \ref{plot3} we
present the plot of the dark energy EoS parameter $\omega_{DE}(z)$
for the $k$-Essence $f(R)$ gravity (blue curve) and for the
power-law corrected $R^2$ model (red curve) as functions of the
redshift. In the plot we can see clearly that in the power-law
corrected $R^2$ model case (red curve) the oscillations are
strongly pronounced as the redshift increases, and in contrast, in
the $k$-Essence $f(R)$ gravity case, the oscillations are
completely absent. Also it is notable that in the $k$-Essence
$f(R)$ gravity case, the dark energy EoS parameter is slowly
varying, with $\omega_{DE}(10)=-0.99967$, while as we mentioned,
the value at redshift zero is $\omega_{DE}=-0.999667$. In Table
\ref{table1} we gathered all the results for several statefinder
and cosmological quantities, for both the $k$-Essence $f(R)$
gravity and power-law corrected $R^2$ models, and also the Planck
and the $\Lambda$CDM model values. Another important cosmological
quantity is the dark energy density parameter
$\Omega_{DE}(z)=\frac{\rho_{DE}}{\rho_{tot}}$, which in terms of
$Y_H$ is given below,
\begin{equation}\label{omegaglarge}
\Omega_{DE}(z)=\frac{Y_H(z)}{Y_H(z)+(z+1)^3+\chi (z+1)^4}\, .
\end{equation}
The value of $\Omega_{DE}$ at present time, for the $k$-Essence
$f(R)$ gravity model is $\Omega_{DE}(0)=0.679553$ which is
compatible with the latest Planck value $\Omega_{DE}=0.6847\pm
0.0073$. Note also that the power-law corrected $R^2$ model yields
$\Omega_{DE}(0)=0.681369$. Let us now consider the behavior of
several well-known statefinder quantities, and we shall be
interested in the deceleration parameter $q$, the jerk parameter
$j$, the parameter $Om(z)$ \cite{Sahni:2014ooa} and finally the
parameter $s$ \cite{Sahni:2002fz}, which are given below,
\begin{align}\label{statefinders}
& q=-1-\frac{\dot{H}}{H^2}\, , \,\,\,
j=\frac{\ddot{H}}{H^3}-3q-2\, , \\ \notag &
s=\frac{j-1}{3(q-\frac{1}{2})}\, , \,\,\,
Om(z)=\frac{\frac{H(z)^2}{H_0^2}-1}{(1+z)^3-1}\, .
\end{align}
All the statefinder quantities are valuable for the study of the
dark energy era, since they depend solely on the Hubble rate and
its higher derivatives, hence they depend explicitly on the
geometry of spacetime via the Hubble rate. The values of the
aforementioned statefinder quantities for the $\Lambda$CDM model
are presented in Table \ref{table1} where we also present the
corresponding values for the $k$-Essence $f(R)$ gravity model and
for the power-law corrected $R^2$ model, for several redshifts.
Let us firstly consider the statefinder $Om(z)$ and in Fig.
\ref{plot4} we present the plots of $Om(z)$ for the $k$-Essence
$f(R)$ gravity (blue curve), for the power-law corrected $R^2$
model (red curve) and the $\Lambda$CDM model (black dashed curve)
as functions of the redshift. As it can be seen, for this specific
statefinder quantity, there are differences between the three
models, and also no oscillations are observed in the $f(R)$
gravity related models, as expected since $Om(z)$ depends only on
the Hubble rate and not on its derivatives. Accordingly in Fig.
\ref{plot4a} we present the plot of the deceleration parameter $q$
as a function of the redshift, for the $k$-Essence $f(R)$ gravity
(blue curve), for the power-law corrected $R^2$ model (red curve).
In this case, the oscillations in the $k$-Essence $f(R)$ gravity
model are completely eliminated, while these are present for the
power-law corrected $R^2$ model. It is notable that both models
are almost indistinguishable from the $\Lambda$CDM model.
Finally, in Fig. \ref{plot5} we present the plots of the
statefinder function jerk $j$ (left plot) and the of $s$ (right
plot) for the $k$-Essence $f(R)$ gravity (blue curves), for the
power-law corrected $R^2$ model (red curves). In this case too,
the oscillations are completely absent in the $k$-Essence $f(R)$
gravity case. Also, we need to note that the jerk for the
$k$-Essence $f(R)$ gravity case is almost indistinguishable from
the $\Lambda$CDM value, however it is not constant, as is probably
inferred from Fig. \ref{plot5}, it is slowly varying though. For
example its value at a redshift $z=10$ is $j(10)=1.00677$ while at
$z=0$ is $j(0)=0.99952$, which are both very close to the
$\Lambda$CDM value $j=1$. In Table \ref{table1} we gather the
values of several cosmological quantities and statefinders for
various redshifts values, for the $k$-Essence $f(R)$ gravity and
the power-law corrected $R^2$ models, and we also quote the
corresponding $\Lambda$CDM values, the latest Planck constraints
or SNe Ia constraints applying for the deceleration parameter. All
the models are viable, however the $k$-Essence $f(R)$ gravity
model seems to be more close to the $\Lambda$CDM model for most of
the quantities considered.
\begin{table}[h!]
  \begin{center}
    \caption{\emph{\textbf{Cosmological Parameters Values for the $k$-Essence $f(R)$ Gravity Model, the power-law corrected $R^2$ model, the $\Lambda$CDM Model and the Planck 2018 data where available (quoting only the Planck 2018 constraint if available ).}}}
    \label{table1}
    \begin{tabular}{r|r|r|r|r}
     \hline
      \textbf{Cosmological Parameter} & \textbf{Odintsov-Oikonomou $f(R)$ } & \textbf{$k$-Essence $f(R)$ } & \textbf{$\Lambda$CDM} $\,\,\,\,$ &  \textbf{Planck 2018 or SNe IA}\footnote{For the deceleration parameter only, based on Ref. \cite{Kumar:2011sw} } \\

           $\,\,\,\,\,\,\,\,\,\,$ &  \textbf{Gravity Model Value} & \textbf{ Gravity Value}  & $\,\,\,\,\,\,\,\,\,\,$ & $\,\,\,\,\,\,\,\,\,\,$  \\
           \hline
      $\Omega_{DE}(0)$ & 0.683948 & 0.679553 & - & $0.6847\pm 0.0073$
      \\  \hline
      $\omega_{DE}(0)$ & -0.995205 & -0.999667 & - & $-1.018\pm 0.031$
      \\  \hline
      $Om(0.000000001)$ & 0.319364 & 0.320707 & - & $0.3153\pm
      0.007$\\  \hline
      $q(0)$ & -0.520954 & -0.51894 & -0.535 & $-0.38\pm 0.05$(SNe Ia data)  \cite{Kumar:2011sw}
      \\ \hline
      $j(0)$ & 1.00319 & 0.99952 & 1 & -\\  \hline
      $j(5)$ & 1.19873 & 1.00362 & 1 & -\\  \hline
      $j(10)$ & 4.25181 & 1.00677 & 1 &- \\  \hline
      $s(0)$ & -0.00104169 & 0.00015711 & 0 & -\\  \hline
    \end{tabular}
  \end{center}
\end{table}
In conclusion, our results indicate that the effect of the
$k$-Essence terms on the late-time phenomenology of $f(R)$ gravity
is very specific, and particularly, it completely eliminates the
dark energy oscillations. These dark energy oscillations were
present for the power-law corrected $R^2$ model especially at
large redshifts, and were more pronounced for quantities that
contained higher derivatives of the Hubble rate. As we
demonstrated though, the $k$-Essence terms utterly change the
picture, by completely eliminating the oscillations and more
importantly they also provide a better fit to the $\Lambda$CDM
model for most of the statefinder quantities we considered,
especially the jerk and the deceleration parameter.

Another issue that is worth mentioning is the effect of initial
conditions of the scalar field on the whole numerical study. It
seems that the same picture occurs for a wide range of initial
conditions values, for example similar results are obtained if we
choose,
\begin{equation}\label{initialconditionsscalarfieldfff}
\phi(z=z_{f})= 10^{-2} M_{p}\, ,
\,\,\,\frac{d\phi}{dz}\Big{|}_{z=z_{f}}= -10^{-10} M_{p}\, ,
\end{equation}
but it notable that a stiff system is obtained if
$\phi'(z)\Big{|}_{z=10}\sim -M_p$. Another important issue worthy
of mentioning is the combined effect that possibly the $R^2$ and
the $k$-Essence terms have. Particularly, the $R^2$ is known to
eliminate the dark energy singularities
\cite{Bamba:2008ut,Appleby:2009uf} and refines in general the
behavior of terms that contain higher derivatives of the Hubble
rate, so perhaps the combined effect of the $k$-Essence terms with
the $R^2$ term eliminates completely the dark energy oscillations.
Let us note that the dark energy singularities are connected with
non-linear oscillations of the curvature scalar, during which a
finite-time (sudden) singularity occurs in the curvature. For more
details on this issue and on the way that the $R^2$ term cures the
singularities, we refer the reader to
\cite{Bamba:2008ut,Appleby:2009uf}.

Finally, the parameters that strongly affect the late-time
phenomenology are $f_1$ and $\delta$, with $f_1$ being the
coefficient of the higher order kinetic term in the Lagrangian
$X^2$, and $\delta $ is the exponent of $\sim R^{\delta}$ in the
$f(R)$ gravity of Eq. (\ref{starobinsky}). In fact, for a specific
range of values of these two parameters, the late-time
phenomenology of the $k$-Essence $f(R)$ gravity dramatically
changes, and interesting physical results are obtained. In the
next section we study in brief a phenomenologically interesting
situation.

\begin{figure}[h!]
\centering
\includegraphics[width=20pc]{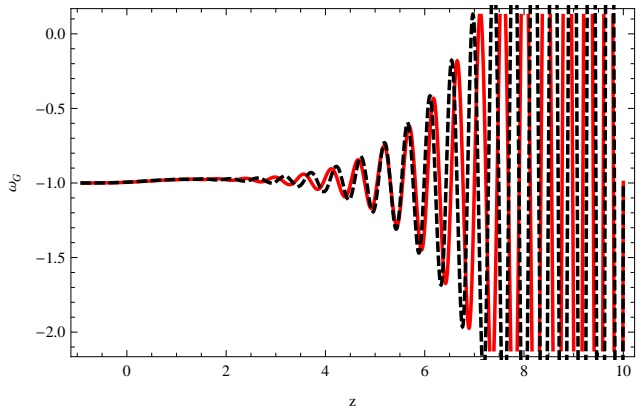}
\caption{Comparison of the redshift dependence of $\omega_{DE}(z)$
for the power-law corrected $R^2$ model, for $H_0=67.4km/sec/Mpc$
(red curve) and $H_0\sim 72km/sec/Mpc$ (black dashed curve), for
the same values of the parameters used in the text.}
\label{refplotfinal}
\end{figure}

Let us note here that we performed all the above calculations by
using the Planck 2018 value for the Hubble rate, namely the one
appearing in Eq. (\ref{H0today}), that is $H_0=67.4km/sec/Mpc$ or
equivalently $H_0=1.37187\times 10^{-33}$eV, but in principle one
could use the value predicted by other sources different from the
CMB, like the ones in Refs.
\cite{Aylor:2018drw,Wong:2019kwg,Verde:2019ivm,Knox:2019rjx,Riess:2016jrr,Migkas:2017vir,Ramos-Ceja:2019zxt},
which predict a tension in the value of $H_0\sim 72km/sec/Mpc$. If
we use the value $H_0\sim 72km/sec/Mpc$, the whole analysis we
performed in this section shall change as it is conceivable, and
in order for a correct viable late-time phenomenology to be
obtained, the values of the free parameters must change. For
example, if we use the value $H_0\sim 72km/sec/Mpc$ or
equivalently $H_0=2.13512\times 10^{-33}$eV, the plots we
presented in this section will change. In Fig. \ref{refplotfinal}
we present the comparison of the redshift dependence of
$\omega_{DE}(z)$ for the power-law corrected $R^2$ model, for
$H_0=67.4km/sec/Mpc$ (red curve) and $H_0\sim 72km/sec/Mpc$ (black
dashed curve), for the same values of the parameters used in the
text of this section. As it can be seen, there are some changes,
and this will be more apparent if we evaluate the value of
$\omega(0)$, which by choosing $H_0\sim 72km/sec/Mpc$ we get
$\omega_{DE}(0)\simeq -0.994612$, which is slightly different from
the value found in Table \ref{table1} for $H_0\sim
67.4km/sec/Mpc$, namely $\omega_{DE}(0)=-0.995205$. The slight
difference is caused by the change in the parameter
$m_s^2=\frac{\kappa^2\rho^{(0)}_m}{3}=H_0\Omega_m$, which for the
value $H_0=67.4km/sec/Mpc$ is equal to $m_s^2=1.87101\times
10^{-67}$eV$^2$ while when the value $H_0\sim 72km/sec/Mpc$ is
used, the value of $m_s^2$ becomes $m_s^2=2.13512\times
10^{-67}$eV$^2$. The same applies for the $k$-Essence $f(R)$
gravity theory, but we omit the details for brevity.

Before closing, it is worth recalling the choices we assigned to
the free parameters of model, which is important in order to
understand how much constrained is the model. Essentially, the
parameters that are free, without any direct correlation with the
observational data, are $\gamma$, $\delta$, $\Lambda$ appearing in
the $f(R)$ gravity model of Eq. (\ref{starobinsky}), and the
parameters $f_1$ and $m$ related to the coupling and the exponent
of the higher order kinetic term $\sim f_1 X^m$ in Eq. Eq.
(\ref{ft17}). The choice of $m$ we assumed in the text, was $m=2$,
and this was for simplicity reasons, because the quadratic higher
order term is the simplest case we can have. Now the parameter
$\Lambda$ was chosen $\Lambda\simeq 11.895\times 10^{-67}$eV$^2$
for two reasons, firstly in order for it to be of the order of the
present day cosmological constant, and secondly in order the fine
tuning on this parameter leads to more aesthetically optimal
values for the parameters $\gamma$ and $\delta$, and more
importantly all the choices lead to a viable cosmology. The
parameter $\delta$ must be in the interval $0<\delta <1$, so if we
choose $\Lambda\simeq 11.895\times 10^{-67}$eV$^2$, the values
$\gamma=2$ and $\delta=1/100$ result to a phenomenologically
viable late-time phenomenology \cite{Odintsov:2020nwm}. A slight
change in $\Lambda$ might require not so aesthetically appealing
values for $\gamma$ and $\delta$ in order to achieve a viable
late-time phenomenology, for instance $\gamma=0.561$ and
$\delta=1/103$ for $\Lambda\simeq 7\times 10^{-67}$eV$^2$. Now the
parameter $f_1$ is the only one that requires fine tuning, in
order to obtain a viable late-time phenomenology for the
$k$-Essence $f(R)$ gravity model, and the choice was $f_{1}\sim
\Lambda^{2-2m}$ appearing in Eq. (\ref{ft26}).

\section{Dynamically Screened Dark Energy Era at Low Redshifts}

One of the latest observations in the last decade was the
measurement of the Hubble rate at low redshift $z\sim 2.34$, with
value $H(z=2.34)=222km/Mpc/sec$ \cite{Delubac:2014aqe}. It must be
mentioned the result of \cite{Delubac:2014aqe}, which indicates
that the Hubble rate at higher redshift increases, is also
supported by other groups in the literature, see for example
\cite{Moresco:2016mzx,Guo:2015gpa,Stern:2009ep,Chuang:2012qt},
however, other measurements at the same redshift do not exist to
our knowledge, so caution is needed. For the purposes of this
article, we shall assume that the measurement of
\cite{Delubac:2014aqe} is correct, but in principle, in order to
consider this result legitimate, this value has to be confirmed.

\begin{figure}[h!]
\centering
\includegraphics[width=20pc]{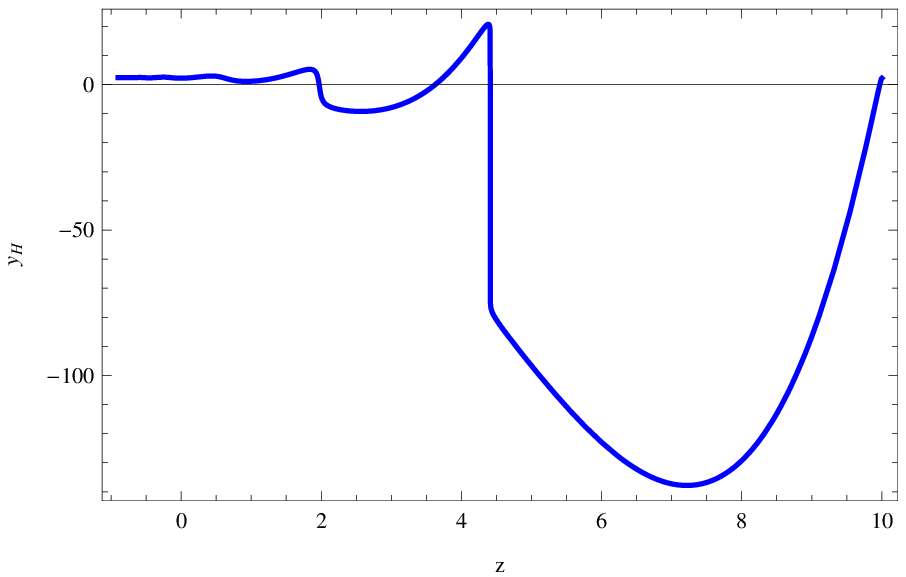}
\includegraphics[width=20pc]{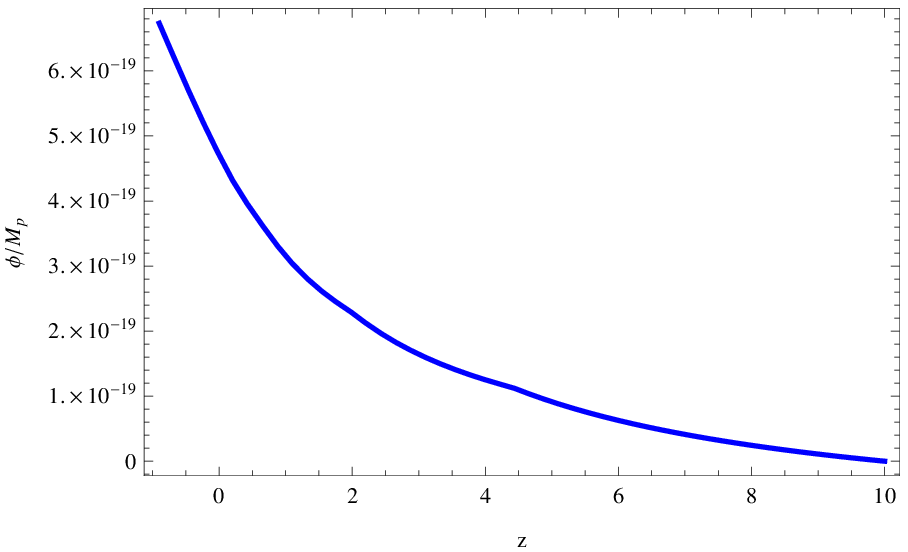}
\caption{Plot of the statefinder function $Y_H$ (left plot) and
the of $\phi(z)/M_p$ (right plot) for the compensating $k$-Essence
$f(R)$ gravity.} \label{plot6}
\end{figure}

\begin{figure}[h!]
\centering
\includegraphics[width=20pc]{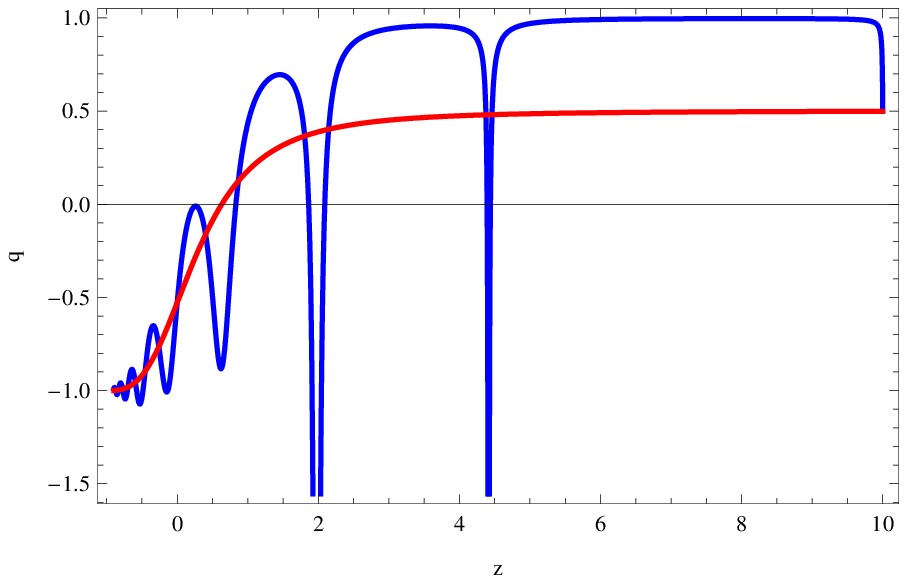}
\includegraphics[width=20pc]{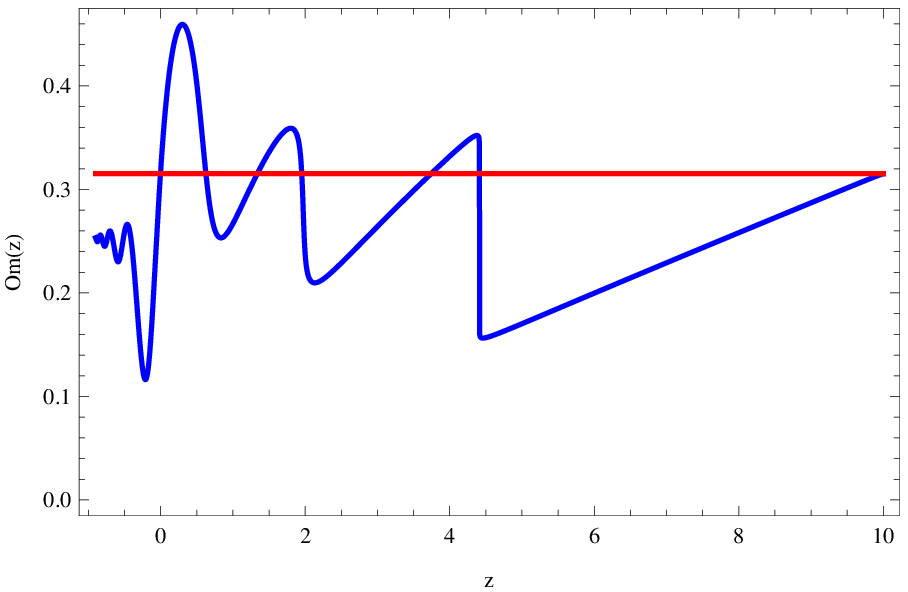}
\caption{Plots of the deceleration parameter $q$ (left plot) and
the $Om(z)$ statefinder (right plot), for the compensating
$k$-Essence $f(R)$ gravity model (blue curves) and the
$\Lambda$CDM model (red curves).} \label{plot7}
\end{figure}

There are two ways to interpret the result of
\cite{Delubac:2014aqe}, if it is assumed to be correct, one to
substitute $\rho_{DE}=0$ in the Friedmann equation (\ref{ft14}),
which would imply that $\Omega_{m}h^2=0.142$ which is obviously in
conflict with the CMB value reported by the Planck data $\Omega_c
h^2=0.12\pm 0.001$ \cite{Aghanim:2018eyx}. Obviously, this would
be a curious result, in the absence of dark energy, so the second
way to interpret it would be to assume that dark energy terms are
present, and these would result to negative $\rho_{DE}$ at $z\sim
2.34$ \cite{Sahni:2014ooa}. It is not the first time that negative
dark energy density appears in the literature, see for example
Refs. \cite{Ahmed:2002mj,Cardenas:2014jya} and references therein.
What we would like to briefly demonstrate in this subsection is
the possibility to generate a negative $\rho_{DE}$ contribution
for redshifts $z\sim 2$, without the need of introducing a
compensating dark energy term by hand, as it is done in Ref.
\cite{Sahni:2014ooa}. In fact, the $k$-Essence $f(R)$ gravity
framework generates such a phenomenological behavior, by simply
choosing appropriately the parameters of the model. We shall again
consider the $k$-Essence $f(R)$ gravity model of the previous
section, with the same conventions for the cosmological
parameters, and the same initial conditions, with the difference
that we choose $f_1=4.01\times 10^{-47}\Lambda^{2-2m}$ and also
the power of the  $\sim R^{\delta}$ term to be $\delta =1/15$.
Now, a negative $\rho_{DE}$ contribution in the context of our
work, would imply negative values of the statefinder function
$Y_H(z)$, and this is the aim of this subsection, to demonstrate
that this is possible by using the $k$-Essence $f(R)$ gravity
theory framework. We numerically solved the Friedmann equation,
and the results of the behavior of the statefinder function
$Y_H(z)$ as a function of the redshift are presented in the left
plot of Fig. \ref{plot6}, while in the right plot we present the
behavior of $\phi(z)/M_p$. As it can be seen, at $z\sim 2-3.8$ the
function $Y_H(z)$ develops negative values, and as the redshift
decreases, $Y_H(z)$ increases until present time. In Table
\ref{table2} we quote the values of the statefinder $Y_H(z)$ for
various redshifts for both the compensating $k$-Essence $f(R)$
gravity and for the power-law corrected $R^2$ model of the
previous section. As it can be seen, for redshifts $z>2$ the
differences are quite significant.
\begin{table}[h!]
  \begin{center}
    \caption{\emph{\textbf{Values of the $Y_H(z)$ for both the compensating and power-law corrected $R^2$ model.}}}
    \label{table2}
    \begin{tabular}{r|r|r}
     \hline
      \textbf{$Y_H(z)$} & \textbf{$k$-Essence $f(R)$ Gravity} & \textbf{power-law corrected $R^2$ model Value}  \\
           \hline
      $Y_H(0)$ & 2.18372 & 2.16471 \\
      $Y_H(0.5)$ & 2.85154 & 2.18795\\
      $Y_H(1)$ & 1.1292 & 2.22296 \\
      $Y_H(1.5)$ & 3.19358 & 2.25955 \\
      $Y_H(1.969)$ & 0.0617874 & 2.28988\\
      $Y_H(2)$ & -4.18054 & 2.29146 \\
       $Y_H(2.34)$ & -8.89681 & 2.30763 \\
     $Y_H(3.7)$ & 1.49666 & 2.36204 \\
      \hline
    \end{tabular}
  \end{center}
\end{table}
Thus the $k$-Essence $f(R)$ gravity theoretical framework provides
a natural compensating dark energy mechanism, which we do not
introduce by hand. We shall call it for the purposes of this paper
compensating $k$-Essence $f(R)$ gravity.
It is worth investigating further the phenomenology of this case,
so we evaluated the dark energy EoS parameter $\omega_{DE}(0)$ at
present time and also the dark energy density parameter
$\Omega_{DE}$, and the results are,
\begin{equation}\label{rescomp}
\Omega_{DE}(0)=0.685836\, ,\,\,\, \omega_{DE}(0)=-0.995346\, ,
\end{equation}
which are both compatible with the latest Planck constraints
$\Omega_{DE}=0.6847\pm 0.0073$ and $\omega_{DE}=-1.018\pm 0.031$.
In general, by analyzing the statefinder quantities it turns out
that the resulting phenomenology is marginally appealing, though
peculiar and quite different from the case studied in the previous
section. In order to see this more clearly, we chose to present in
Fig. \ref{plot7} the plots of the deceleration parameter $q$ (left
plot) and the $Om(z)$ statefinder (right plot), for the
compensating $k$-Essence $f(R)$ gravity model (blue curves) and
the $\Lambda$CDM model (red curves). As it is obvious from the
plots, the compensating $k$-Essence $f(R)$ gravity model is quite
different from the $\Lambda$CDM model, and only at very small
redshifts near the present-time era there is some overlap between
the two models, at least when the deceleration parameter is
considered. It is worth quoting here the values of the
deceleration parameter and $Om(z)$ statefinder for $z=0$ for the
compensating $k$-Essence $f(R)$ gravity, and these are,
\begin{equation}\label{finaleqn}
q(0)=-0.523917\, , \,\,\,Om(0.000000001)=0.317389\, ,
\end{equation}
which are quite close to the $\Lambda$CDM values $q=-0.535$ and
$Om(z)=0.3153$. We refrain from going into further details, since
the general picture is obvious, the result is that we obtain a
marginally compatible to the $\Lambda$CDM phenomenology, only at
low redshifts, while at larger redshifts, there are differences.
However, with the compensating $k$-Essence $f(R)$ gravity model we
obtain negative dark energy density at nearly $z\sim 2.34$, which
can explain the observational values of the Hubble rate
\cite{Delubac:2014aqe} at the same redshift. Our model presented
in this section offers a phenomenological description which can
provide phenomenologically acceptable values to some of the
cosmological quantities of interest at redshift zero (present
day), like the $\Omega_{DE}(0)$, $\omega_{DE}(0)$, the
deceleration parameter and the statefinder $Om(z)$, and also can
yield negative values for the statefinder $Y_H(z)$ at redshift
$z\sim 2.34$. However, the model cannot be considered a fully
correct description of the Universe, neither at present time, nor
at higher redshifts, since the overall behavior of some observable
quantities is not phenomenologically acceptable.
\begin{figure}[h!]
\centering
\includegraphics[width=20pc]{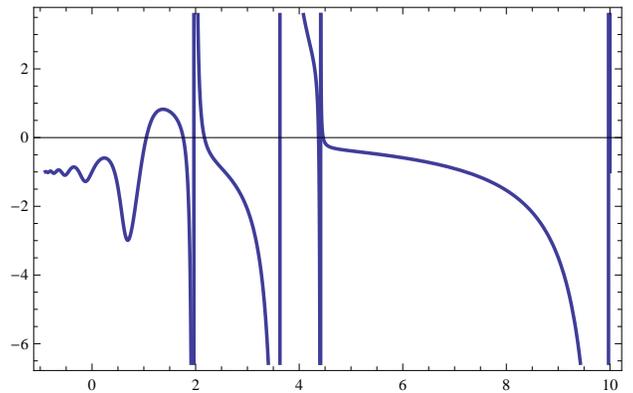}
\caption{Plot of $\omega_{DE}(z)$ for the compensating $k$-Essence
$f(R)$ gravity model for $z=[0,10]$.} \label{plotref1}
\end{figure}
For example, in Fig. \ref{plotref1} we plot $\omega_{DE}(z)$ for
redshifts $z=[0,10]$ for the compensating $k$-Essence $f(R)$
gravity model, and the result is rather unappealing and
phenomenologically not acceptable, although the value of
$\omega_{DE}(z)$ at redshift zero is phenomenologically
acceptable. Also the blue curves in Fig. \ref{plotref1} do not
provide an optimal fit to the $\Lambda$CDM model. What we aimed in
this section is to demonstrate that negative values of the
statefinder $Y_H(z)$ can be obtained by the $k$-Essence $f(R)$
gravity model, by appropriately tuning some parameters of the
model, and at the same time obtaining compatibility with the
observational data for some of the observable quantities at
present time. Our description though cannot be considered as a
fully viable description of the Universe, and also the results of
Ref. \cite{Delubac:2014aqe} must also be widely accepted in order
to further study how our model can be a viable description of the
Universe, up to redshifts $z\sim 3$.

\section{Conclusions}

In this paper we studied the effects of $k$-Essence terms in the
late-time phenomenology of $f(R)$ gravity in the presence of cold
dark matter and radiation perfect fluids. We chose an $f(R)$
gravity model which has quite phenomenologically appealing
late-time properties, and we assumed that a canonical scalar field
term and a higher order kinetic term are also present in the
gravitational Lagrangian. With regard to the higher order kinetic
term, we studied the case that this term is a quadratic term of
the form $\sim f_1 X^2$. The dimensionful parameter plays an
important role on the phenomenology as it turns, and in fact, this
term in conjunction with an $R^{\delta}$ term appearing in the
$f(R)$ function, crucially affect the late-time phenomenology of
the $k$-Essence $f(R)$ gravity model. The power-law corrected
$R^2$ model apart from the fact that it is similar to the
$\Lambda$CDM model, it was plagued with the issue of dark energy
oscillations at large redshifts. These dark energy oscillations
are more pronounced when physical quantities that contain higher
derivatives of the Hubble rate are considered. Our aim in this
work was to investigate whether these dark energy oscillations are
eliminated from the late-time era in the context of $k$-Essence
$f(R)$ gravity, and as it turns, this occurs for a wide range of
initial conditions imposed on the scalar field, and for some
values of the parameter $f_1$ which is the coefficient of the
quadratic kinetic term $X^2$. As we demonstrated, a viable
late-time phenomenology is produced by the $k$-Essence $f(R)$
gravity, without the presence of dark energy oscillations, at
least up to a redshift $z\sim 10$. We studied several cosmological
quantities of cosmological interest, such as the dark energy EoS
parameter, the dark energy density parameter, and several
statefinder quantities. In all the cases, the $k$-Essence $f(R)$
gravity model was almost indistinguishable from the $\Lambda$CDM
model, as for example in the case of the deceleration parameter
$q$ and the jerk, and in all cases the dark energy oscillations
were absent. We also compared directly the power-law corrected
$R^2$ model and the $k$-Essence $f(R)$ gravity models, to see the
difference between the two models and the complete absence of dark
energy oscillations for the $k$-Essence $f(R)$ gravity model.
Apart from this major issue, we also investigated how the
$k$-Essence $f(R)$ gravity model could explain the 2014
observational data concerning the Hubble rate at redshift $z\sim
2.34$. In the presence of dark energy, this observation would
require a negative energy density for dark energy. As we
demonstrated, by using a specific set of values for the parameter
$f_1$ and the exponent of $R^{\delta}$, we achieved a negative
energy density for redshifts $z\sim 2-3.8$, and also we
demonstrated that the resulting values of the dark energy EoS
parameter and of the dark energy density parameter were compatible
with the latest Planck data, however the model produced quite
different behavior of the statefinder parameters in comparison to
the $\Lambda$CDM model. Our findings support the idea that the
$f(R)$ gravity theory and its modifications, viewed as a perfect
fluid \cite{Capozziello:2018ddp}, can mimic a
curvature-quintessence like behavior \cite{Capozziello:2002rd}, at
least at late-times. A good question though is how does the
$k$-essence term affects the large scale structure formation
during the matter domination era. This question is non-trivial to
answer in brief though, and a focused simulation on this issue
should be performed.

Finally, let us note that in our study we did not take into
account the presence of a potential for the scalar field, and we
did not investigate at all the case that the scalar field is a
phantom scalar. We hope to address these issues in a future work.

\section*{Acknowledgments}

This work is supported by MINECO (Spain), FIS2016-76363-P, and by
project 2017 SGR247 (AGAUR, Catalonia) (S.D.O).

\end{document}